\documentclass[reprint,aip,rsi,graphicx]{revtex4-1}

\usepackage{graphicx}
\usepackage[USenglish]{babel}

\newcommand{\fref}[1]{\mbox{Fig. \ref{#1}}}
\newcommand{\eref}[1]{\mbox{Eq. \ref{#1}}}
\newcommand{\sref}[1]{\mbox{Sec. \ref{#1}}}

\newcommand{\wo}{\omega_0}

\begin{document}

\title{Compact RF resonator for cryogenic ion traps}

\author{D. Gandolfi} \affiliation{Dipartimento di Fisica - Universit\`a degli Studi di Trento, I 38123 Trento, Italy} \affiliation{Institut f\"ur Experimentalphysik - Universit\"at Innsbruck, A 6020 Innsbruck, Austria}

\author{M. Niedermayr} \affiliation{Institut f\"ur Experimentalphysik - Universit\"at Innsbruck, A 6020 Innsbruck, Austria}

\author{M. Kumph} \affiliation{Institut f\"ur Experimentalphysik - Universit\"at Innsbruck, A 6020 Innsbruck, Austria}

\author{M. Brownnutt} \email[Correspondence should be addressed to ]{Michael.Brownnutt@uibk.ac.at} \affiliation{Institut f\"ur Experimentalphysik - Universit\"at Innsbruck, A 6020 Innsbruck, Austria}

\author{R. Blatt} \affiliation{Institut f\"ur Experimentalphysik - Universit\"at Innsbruck, A 6020 Innsbruck, Austria}  \affiliation{Institut f\"ur Quantenoptik und Quanteninformation - \"Osterreichische Akademie der Wissenschaften, A 6020 Innsbruck, Austria}

\date{\today}

\begin{abstract}
We report on the investigation and implementation of a lumped-component, radio-frequency resonator used in a cryogenic vacuum environment to drive an ion trap. The resonator was required to achieve the voltages necessary to trap ($\sim100\,$V), while dissipating as little power as possible ($\lesssim 250\,$mW). Ultimately a voltage gain of 100 was measured at 5.7 K. Single $^{40}\rm{Ca}^{+}$ ions were confined in a trap driven by this device, providing proof of successful resonator operation at low temperature.
\end{abstract}


\maketitle

\section{Introduction}
\label{sec:introduction}
In the last decade trapped ions have proved to be a suitable system to perform quantum computation (QC) and quantum information processing (QIP) \cite{blatt2}. Proof-of-principle demonstrations of several algorithms were performed\cite{gulde2,chiaverini1,brickman,lanyon2011}, though the complete control of a large number of qubits is still an outstanding challenge\cite{roadmap,steane2}. One proposed way to scale-up the number of qubits, is the  use of a multiplexed array of small traps, with segmented DC electrodes used to shuttle ions from the memory region to the processing region\cite{kielpinski,schaetz}. A reasonable way to build these arrays of traps is using planar surface traps \cite{chiaverini2,seidelin}, which can benefit from photolithographic techniques to achieve miniaturization to micrometer feature sizes. However, as the ion-electrode distance is reduced, ion heating becomes significant\cite{turchette}. This can be mitigated to some extent by operating at cryogenic temperatures\cite{
labaziewicz1}.

Despite great progress in the miniaturization of ion-traps, little has changed regarding the trap-drive electronics. In order to limit Joule dissipation and RF crosstalk, the relatively high radio-frequency voltages necessary for trapping, typically around a hundred volts, can not be applied through long cables in the cryostat (see \fref{fig:circuit}(a)). At the same time it is not possible to shorten the cables, since the thermal load from the room temperature part would be too big. The ideal solution is to send a small amount of RF power through the (long) cables, and amplify the voltage with a resonator as close as possible to the trap. The common choice adopted for ion trapping is the use of bulky helical resonators \cite{macalpine}, which can also be operated at cryogenic temperatures \cite{poitzsch}. 

In this work the resonators are scaled to much smaller sizes and made using lumped components, namely series RLC resonators. These circuits are naturally compact, and the small volume makes them suitable for the use in a cryostat, or when several resonators should be used to drive different RF-electrodes\cite{kumph}. In this article we show that these small resonators can be effectively used to trap ions, while dissipating a negligible amount of RF-power. The resonator described here reached a voltage step-up as high as $100$ at low temperature.

\section{Overview of the experiment}
\label{sec:overview}
\begin{figure}
 \includegraphics{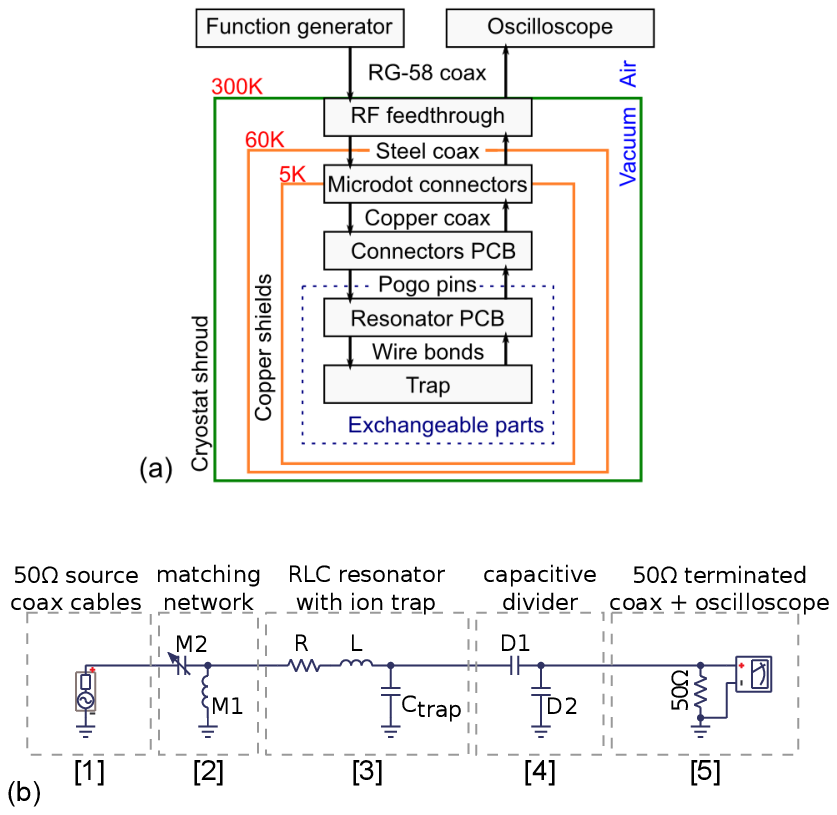}
 \caption{\label{fig:circuit}(a) Complete schematic of the trap-driving electronics. The approximate temperatures in each stage are also shown. (b) Detailed schematic of the resonator and of the auxiliary electronics. The radio-frequency source is a commercial function generator. The matched RLC resonator provides the necessary voltage step-up while the voltage divider and the oscilloscope are used for the measurement and the voltage control.}
\end{figure}

An overview of the experiment is shown in \fref{fig:circuit}. The RF voltage source is a commercial function generator (\emph{Thurlby Thandar Instruments TG 4001}), with an output impedance of $50\,\Omega$. The RF signal is applied, via coaxial cables, to the resonator - and then to the trap - inside the cryostat (\emph{ARS CS210S-GMX-20}). The transmission line is composed of three different types of coaxial cables and two types of RF connector. The first cable is a standard copper-core RG-58. The second cable (\emph{Lakeshore SS-32}) is made from steel for thermal decoupling from the room-temperature part of the setup to the cryogenic-temperature part. The steel's relatively high resistance accounts for a measured attenuation of $3.2\,$dB for frequencies of about $10\,$MHz. Moreover, the cable characteristic impedance is $40\,\Omega$ and it is not well-matched to the rest of the transmission line. However, the calculated return loss due to this mismatch is negligible ($\sim0.1\,\%$). The last cable is a 
very thin and flexible coaxial cable (\emph{Samtec MH081}). The core material is silver-plated copper and the characteristic impedance is $50\,\Omega$. The first connector is the coaxial feedthrough (\emph{Accu-Glass 25D-5CX2-450}), mounted on the outer shroud of the cryostat, while the second is an array of microdot straight connectors (\emph{Tyco Electronics 141-0001-0001} and \emph{142-0000-0001}), anchored on the $5\,$K copper shield. The microdot connectors showed excellent performance at low temperature, even after several cooling cycles.
After this coaxial transmission line the RF voltage is applied -- through soldered connections -- to a PCB board, where gold-plated pogo-pins are soldered (\emph{Mill-Max 0929-0-15-20-75-14-11-0}). It was decided to use these connectors in order to have the possibility to easily interchange the last part of the circuit, where the resonator and the trap are, without major changes to the setup.

The exchangeable part of the circuit is the matched RLC resonator with the ion trap. The matching circuit (part [2] in \fref{fig:circuit}(b)) is a simple ``L-section'' network\cite{abrie}, because the resonator is meant to be driven within a narrow range of frequencies. The series RLC resonator (\fref{fig:circuit}(b) [3]) is actually composed of only an inductor and the trap capacitance, but an effective resistance must be used in the model to take into account any dissipation in the resonator. The value of this effective resistance influences the quality of the resonator, so keeping it as low as possible is one of the main concerns. The trap is a planar surface trap, with an ion-electrode distance of $454\,\mu$m and a distance between electrodes of $20\,\mu$m. It is photolithographically patterned, depositing, by means of e-beam evaporation, $2\,$nm of Ti (adhesive layer) and $300\,$nm of Au on a substrate of SiO$_2$ ($500\,\mu$m).

The last two blocks in the schematic (\fref{fig:circuit}(b) [4] and [5]) are the capacitive divider and an oscilloscope (\emph{Tektronix TDS 2004B}). These are used to measure and control the voltage at the resonator's output. The purpose of the capacitive divider is to increase the input impedance of the measurement device (in this case the  oscilloscope plus the transmission line) and to make the capacitance of this ``probe'' independent of the length of the coaxial cable. The divider is connected in parallel to the trap; for this reason its capacitance has to be taken into account when calculating the resonator's characteristics.

\section{Circuit analysis}
\label{sec:analysis}
In this section the non-commercial parts (matching network, resonator and capacitive divider) will be described; the motivations for their necessity and the equations for their characterization are discussed.

\subsection{Matching network}
\label{sec:matching}
Since the impedance of the resonator at resonance is, in general, different from the impedance of the transmission line, in order to reduce reflections of power it is necessary to add a matching network to the circuit. As for the voltage-amplifying circuit, low power consumption and cryogenic-compatible components constrain the matching network to be passive.

The obvious choice for an easily customizable single-frequency passive matching network is the ``L-section'' (a voltage divider built with two reactive components). At resonance the impedance of the resonator will be purely real and, usually, smaller than $50\,\Omega$. This means that the first matching reactance - seen from the resonator's point of view - has to be connected in series. The second component has to be an opposite reactance (a capacitor, if the first was an inductor, or vice-versa) connected in parallel. In both of these two possible configurations, the resonator and the trap do not have a DC path to ground because of the capacitor. In order to remove any DC bias voltage and reduce the low-frequency noise on the RF trap electrodes, which could perturb the ions, a DC path to ground should be provided.

On a closer analysis, the role of the first (series) reactance is simply to shift the resonance frequency until the impedance of the resonator (which is frequency-dependent) can be matched with just a parallel component. An alternative matching, thus, can be achieved operating the resonator at a frequency - usually different from the resonance - where its impedance can be matched with only one parallel reactance. The two frequencies where matching with one parallel reactance is possible are denoted here as $\omega_{\rm{L}}$ (low) and $\omega_{\rm{H}}$ (high). The matching at these two frequencies can be accomplished with an inductor or a capacitor, respectively. If this reactance (M$1$ in \fref{fig:circuit}(b)) is inductive, then the DC path to ground is ensured. A second impedance can then be added (M$2$) in series for fine-tuning the matching, without interrupting the DC path.

\begin{figure}[h]
 \includegraphics{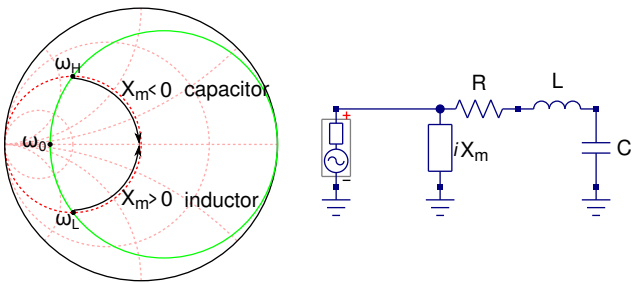}
 \caption{\label{fig:matching_trick}Smith chart of the resonator's impedance as a function of the frequency (continuous green line, frequency increasing clockwise). The two frequencies where matching with one parallel reactance is possible are marked as $\omega_{\rm{L}}$ (low) and $\omega_{\rm{H}}$ (high). The matching can be accomplished with an inductor or a capacitor, respectively.}
\end{figure}

In \fref{fig:matching_trick} the resonator's impedance is plotted as a function of the frequency (continuous green line). It can be shown that the frequency $\omega_{\rm{L}}$ and the matching reactance $X_{\rm{m}}$ can be expressed as
\begin{eqnarray}
\omega_{\rm{L}}&=&\wo\sqrt{1+\frac{\alpha}{4 Q^2}} - \frac{\wo}{2Q}\sqrt{\alpha} \\
X_{\rm{m}}(\omega_{\rm{L}})&=&\frac{Z_{\rm{s}}}{\sqrt{\alpha}}
\end{eqnarray}
where $Z_{\rm{s}}$ is the RF source's impedance, $\wo$ is the resonator's resonant frequency, $Q = \wo L/R$ is the resonator's quality factor and $\alpha=\frac{Z_{\rm{s}}}{R} - 1$. From an experimental point of view, it is easier to find $\omega_{\rm{L}}$ and $X_{\rm{m}}(\omega_{\rm{L}})$ by measuring the resonator's impedance with a network analyzer.

It is important to understand that matching away of resonance $\wo$ does not compromise the voltage gain given by the resonance enhancement. However, there is an important, unavoidable, drawback connected to the matching: if only reactive components are used in the network, and power conservation holds (which means that the quality factor of the matching circuit $Q_{\rm{M}}\gg Q$), these circuits have to satisfy a strict relation between voltage transformation and impedance transformation. Since we want to match the resonator with an inductor, the matching quality factor $Q_{\rm{M}}$ has the same order of magnitude as $Q$. This means that the overall quality factor will be reduced, by as much as a factor of $2$. Nonetheless the following analysis still holds.

Treating the matching network as a two-port device, power conservation requires
\begin{equation}
|{V_{\rm{in}}}^2/Z_{\rm{in}}| =  |P_{\rm{in}}| =  |P_{\rm{out}}| = |{V_{\rm{out}}}^2/Z_{\rm{out}}|
\end{equation}
(with $V_{\rm{in/out}}$ being the voltage and $Z_{\rm{in/out}}$ the impedance seen from their respective port)
or, equivalently,
\begin{equation}
\label{eq:transf_ratio}
  k := |{V_{\rm{out}}}/{V_{\rm{in}}}| = \sqrt{|Z_{\rm{out}}/Z_{\rm{in}}|}\;.
\end{equation}
In all the relevant cases for resonators driving ion traps, where $|Z_{\rm{out}}|<50\,\Omega$,  $k$ is smaller than $1$.
This means that, when the resonator is matched, the voltage at its input is reduced by a factor proportional to the square root of its impedance.
This factor reduces the overall voltage gain given by the resonator; for this reason the matching network must be taken into account in the resonator design. 

\subsection{RLC resonator}
\label{sec:RLC_resonator}
\subsubsection{Resonator analysis}
\label{sec:resonator_analysis}
The RLC resonator is the part of the circuit where the voltage step-up takes place. It is a voltage divider made of an inductor and a capacitor. The resistance is an effective one and it is used to model the power dissipations of realistic components. The input impedance is easy to calculate and can be written as
\begin{equation}
 Z(\omega) = R + i\omega L -  \frac{i}{\omega C}  = R \left[ 1 + {i}{Q} \left(\frac{\omega}{\wo} - \frac{\wo}{\omega} \right)\right]
\label{eq:RLC_impedance}
\end{equation}
where $\wo = 1/\sqrt{LC}$ is the resonator's resonant frequency and $Q = \wo L/R$ is the resonator's quality factor. On resonance $Z$ is purely real and equal to $R$, which, for the relevant cases, is less than $50\,\Omega$.

At resonance, an unmatched resonator has a voltage gain (the value of the transfer function on resonance) which is equal to the quality factor $Q$. 
However, when a matching network is added, the overall (matching and resonator) voltage gain changes by a factor $k$, as defined in \eref{eq:transf_ratio}. 
The voltage gain near the resonance is
\begin{equation}
G_V(\omega)\simeq\frac{kQ}{\left|1+\frac{i Q}{2}\left(\frac{\omega}{\omega_{\rm{L}}}-\frac{\omega_{\rm{L}}}{\omega}\right)\right|}\;.
\label{eq:gain_w}
\end{equation}
The above equation holds if the frequency dependence of the matching network is negligible with respect to the frequency dependence of the resonator's impedance. Matching with a parallel reactance satisfies this condition.  The peak gain appears on resonance, and it is given by
\begin{equation}
G_V(\wo) \simeq kQ = \sqrt{\frac{\wo L Q}{Z_{\rm{s}}}}\;.
\label{eq:gain}
\end{equation}

\subsubsection{Quality factor Q measurement}
\label{sec:Q_measurement}
To fully characterize a resonator, it is important to measure its quality factor $Q$. Measuring $\wo$ is also important, but trivial.
Thinking of the matched resonator as a circuit with one input port and one output port, it is possible to measure the quality factor $Q$ from either port, independently. From \eref{eq:gain_w} it is possible to show that $Q$ can be measured, at the output port, as
\begin{equation}
\label{eq:QfromV}
 Q = \frac{\wo}{\Delta \omega} = 2 \frac{\wo}{\Delta \omega_{V}}\;.
\end{equation}
Here $\Delta \omega_{V}$ is the measured $-3\,$dB full-width voltage gain bandwidth (referenced to the function generator voltage source), which is twice the voltage gain bandwidth $\Delta \omega$, referenced from the input of the bare RLC resonator.

On the other hand, using a network analyzer at the input port, it is possible to measure the scattering parameter $S$ as a function of frequency\cite{abrie}.
If the resonator were to be matched with a transformer, its impedance would become
\begin{equation}
 Z_{\rm{M}}(\omega) = Z_{\rm{s}} \left[ 1 + {i}{Q} \left(\frac{\omega}{\wo} - \frac{\wo}{\omega} \right)\right]
\end{equation}
and the scattering parameter could be calculated as
\begin{equation}
\label{eq:scattering_S}
 |S| = \left|\frac{Z_{\rm{M}}(\omega) - Z_{\rm{s}}}{Z_{\rm{M}}(\omega) + Z_{\rm{s}}}\right| =  \frac{\frac{Q}{2}\left(\frac{\omega}{\wo} - \frac{\wo}{\omega} \right)}{\left| 1 + {i}\frac{Q}{2} \left(\frac{\omega}{\wo} - \frac{\wo}{\omega} \right)\right|}\;.
\end{equation}
From this equation, it is possible to find that the $1/\sqrt2$ full-width $S$-bandwidth ($\Delta \omega_{S}$) can be used to measure the quality factor, using the formula
\begin{equation}
\label{eq:QfromS}
 Q = 2 \frac{\wo}{\Delta \omega_{S}}\;.
\end{equation}
If different kinds of matching are used, $Z_{\rm{M}}(\omega)$ takes different shapes, and it is not possible to find a generally valid expression. However, as long as the frequency dependence of the matching components is negligible, Eqs. \ref{eq:scattering_S} and \ref{eq:QfromS} continue to be valid (replacing $\wo$ with $\omega_{\rm{L}}$).

\subsubsection{Voltage gain optimization}
\label{sec:optimization}
As mentioned in \sref{sec:introduction}, ion traps need relatively high RF voltages. To achieve this goal without dissipating too much RF power it is necessary to have a high voltage gain in the resonator. If the transmission line's characteristic impedance $Z_{\rm{s}}$ and the operating frequency $\wo$ are fixed, it is obvious from \eref{eq:gain} that the gain can be improved by increasing the values of $L$ and $Q$. 

Note that in the chosen model the effective resistance is completely attributed to the inductor. This is justified by the fact that the quality factors of commercial capacitors are one or two orders of magnitude higher than the quality factors of commercial inductors. In this situation, the general equation for the quality factor of two series component $Q^{-1} = Q^{-1}_{\rm{L}} + Q^{-1}_{\rm{C}}$ reduces to $Q\simeq Q_{\rm{L}}$.

Both of the important characteristics to enhance, $Q_{\rm{L}}$ and $L$, are usually reported on inductors' datasheets, and are limited as follows:
\begin{itemize}
 \item[-] $Q_{\rm{L}}$ is limited by the resistance of the conducting material. This resistance is increased by several effects, like the \emph{skin effect}, the \emph{proximity effect}\cite{abrie} and the induced eddy currents in dissipative materials coupled to the inductor. A way to increase $Q_{\rm{L}}$ is to use good conducting materials and big wire diameters;
 \item[-] $L$ is limited mainly by the parasitic capacitance of the inductor, which introduces a lower bound for the operating frequency. This problem can be reduced by using bigger coils with larger spacing between the windings.
\end{itemize}

\subsection{Capacitive divider}
\label{sec:divider}
A capacitive divider is a voltage divider made with two capacitors. In this circuit its purpose is to increase the input impedance of the measuring apparatus, including the transmission line, in order not to affect the resonator during the measurements. Choosing $D1\ll D2$ and $(\omega D1)^{-1} \gg 50\,\Omega$ (with reference to \fref{fig:circuit}(b)), the impedance seen from the resonator is dominated by the impedance of the first capacitor. In this way it does not depend on the length of the transmission line nor the input impedance of the oscilloscope, and the measurement instrumentation can be disconnected without affecting the resonator. The transmission line is connected in parallel to $D2$. The transmission line impedance adds a frequency-dependence to the divider's voltage transformation if the relation $(\omega D2)^{-1} \ll 50\,\Omega$ is not fulfilled (as in our experiment).

\section{Experimental results}
\label{sec:results}
The circuit testing was done in two steps. The first was the calibration and testing of the capacitive divider.
Here, the choice for a temperature-stable and high quality-factor capacitor led to the use of SMD capacitors with mica dielectric, manufactured by \emph{Cornell Dubilier Electronics}. With reference to \fref{fig:circuit}(b), $D1$ was chosen to be $5\,$pF (part no. \textsl{MC08CD050D-F}), while $D2$ was chosen to be $1\,$nF (part no. \textsl{MC22FD102J-F}), so that the voltage transformation ratio was roughly $200$. The divider was calibrated at room and cryogenic temperatures, proving in this way the capacitors' reliability at low temperature. The measurement of the voltage transformation ratio did not show any significant change with the temperature; however, the measured uncertainty between different realizations of the divider was $6\%$ (due to the uncertainty in the capacitances), leading to a systematic error of the same value in the measured voltage after the divider.

\begin{figure}
 \includegraphics{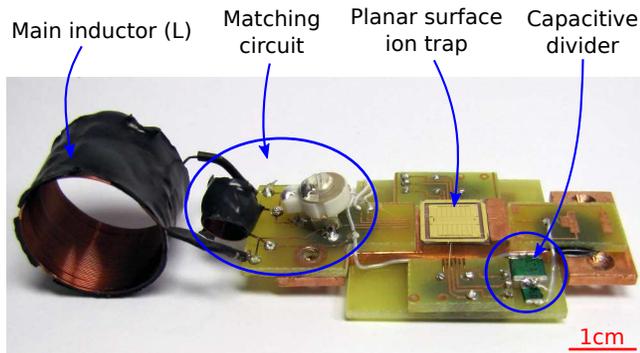}
 \caption{\label{fig:scairy}Final trap-driving circuit prototype. The matched resonator is on the left, the gold-plated planar surface trap is in the center and the capacitive divider on the right. The trap is bound with $25\,\mu$m-gold wires. The prototyping circuit was built on a stack of three FR-4 PCB layers.}
\end{figure}
The second part of the testing was done with the complete circuit. \fref{fig:scairy} shows a picture of the resonator prototype, with the matching network, the capacitive divider and the ion trap in the center. The trap is bonded using $25\,\mu$m-thick gold wires and an ultrasonic wire-bonder.  The circuit is built on a stack of three layers of FR-4 printed circuit boards (PCB).
This material was used for quick development: at low temperature the outgassing rate is low enough to allow trapping of ions. In the circuit, the RF line and the DC lines (for the segmented electrodes) are kept apart, in order to reduce the RF coupling and noise. The resonator and the capacitive divider are placed on two opposite sides of the trap's RF electrode, to measure the trap voltage and to check the continuity of the line at the same time.

The trap capacitance, calculated with finite element software, is $C_{\rm{trap}} \simeq 1\,$pF. The actual value was not measured. 
Several inductors for the resonator were tested. Inductors with cores other than air were limited by dissipation in the core or by cryogenic compatibility issues. The best results for high quality factors were obtained with an air-core inductor manually wound using a copper-coated NbTi wire (\emph{Supercon Inc. 54S43} insulated diameter $0.279\,$mm, superconducting critical temperature $T_{\rm{C}}=9.2\,$K).  As stated before, high $L$ values are limited by the parasitic capacitance in conflict with the required small volume. The physical dimensioning of the inductor can be carried out using some inductance-calculator software\footnote{A good inductance-calculator software can be found at \texttt{http://hamwaves.com/antennas/inductance.html}. It relies on several geometric and empirical correction factors, as given in \cite{RFcoils, inductanceformula, groverinductance}. This applet agreed with our experimental inductance measurements within $10\%$. In addition it can calculate the self-resonance frequency and 
an estimation of the inductor's quality factor.} or formulae\cite{RFcoils, inductanceformula, groverinductance}.
In the final design the inductor has a measured inductance of $45\,\mu$H at $7.56\,$MHz. The inductor's diameter and length are $20\,$mm each. The coil is mechanically stabilized with an epoxy matrix (\emph{Emerson \& Cuming Stycast 2850 FT}).
Since the superconducting material is copper-coated NbTi, the superconducting critical temperature is $9.2\,$K, above the usual minimum working temperature of the cryostat of this experiment ($<6\,$K). The same superconducting cable was used to wind the matching inductor $M1$. $M2$ is a ceramic variable capacitor with range $12-100\,$pF (\emph{Johanson Manufacturing 9328}).

With the circuit mounted inside the cryostat, at room temperature, the matching was tuned by changing $M2$ while checking the scattering parameter $S$ on the input port of the circuit. Closing the vacuum chamber the impedances changed slightly, resulting in a residual reflection $S = 0.027$. The measured resonance frequency was $\wo = 2\pi\cdot7.43\,$MHz, with a voltage gain $G_V = 41\pm2$ and a measured quality factor $Q = 52\pm1$.

\begin{figure}
 \includegraphics{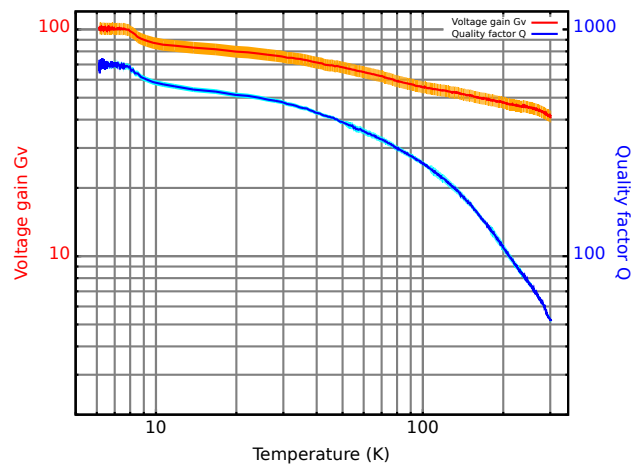}
 \caption{\label{fig:gain_plot}Temperature dependence of the voltage gain $G_V$ and quality factor $Q$ of the trap-drive resonator.
  The error bounds for the voltage gain show a a systematic uncertainty due to the capacitive divider uncertainty, while the instrumental uncertainty is smaller than the line thickness. The error bounds for the quality factor depend on the network analyzer's frequency resolution. The temperature is measured on the copper of the trap carrier.}
\end{figure}

For the cryogenic testing, the temperature was measured on the copper of the trap carrier and in the charcoal of the cryo-pump using two silicon diode temperature sensors (\emph{Lakeshore DT-670-SD-1.4L}). The two measurements agreed within $0.5\,$K. Supported by subsequent measurements, we estimate that when the temperature sensors read $6\,$K, the coil temperature is around $9\,$K. Both the gain and the quality factor were measured as a function of the trap carrier's temperature. The results are reported in \fref{fig:gain_plot}.

The plots of the quality factor $Q$ and voltage gain $G_V$ increase with decreasing temperature, showing an improvement of the resonator at low temperature. At $T = 5.7\,$K (measured at the sensor), the quality factor was $Q = 700\pm30$, one order of magnitude higher than at room temperature. Similarly, the gain was $G_V = 101\pm6$ at $5.7\,$K, more than twice the gain at room temperature.
During the test the resonance frequency showed a small change, moving from $7.43\,$MHz at $300\,$K to $7.641\,$MHz at $5.7\,$K. The scattering parameter $S$ changed from $0.027$ to $0.57$. This is due to the small effective resistance, which at low temperature was $10$ times lower than at room temperature. The matching, which was achieved at $300\,$K, was no longer good at $5.7\,$K.

The voltage gain does not show any evident enhancement when the superconducting-transition temperature is passed. Moreover, a similar realization of the RLC resonator with an inductor built winding an enameled copper wire showed comparable $G_V$ and $Q$.

After the characterization, this circuit was used to successfully trap single $^{40}\rm{Ca}^{+}$ ions.  The output voltage from the function generator was as low as $1.35\,$V which means that the efforts spent to increase the voltage gain enabled trapping of ions using only $18\,$mW of RF power. 

\section{Conclusions}
\label{sec:conclusions}
In conclusion, this article reports the investigation and experimental realization of a lumped component radio-frequency (HF band) resonator used to drive a planar surface ion trap. In contrast to bulky helical resonators, this class of circuits is compact and easily reproducible. Moreover, using tunable components it is possible to change the resonant frequency without major changes in the circuit.

With a carefully built inductor, voltage gains as high as $100$ are achievable at low temperature, and trapping of ions can be achieved with a very small amount of RF power ($18\,$mW for an ion-electrode distance of $454\,\mu$m).

\begin{acknowledgments}
We gratefully acknowledge support from the European Research Council through the project CRYTERION and the Institute for Quantum Information GmbH.
\end{acknowledgments}


\begin{thebibliography}{20}
\makeatletter
\providecommand \@ifxundefined [1]{%
 \@ifx{#1\undefined}
}%
\providecommand \@ifnum [1]{%
 \ifnum #1\expandafter \@firstoftwo
 \else \expandafter \@secondoftwo
 \fi
}%
\providecommand \@ifx [1]{%
 \ifx #1\expandafter \@firstoftwo
 \else \expandafter \@secondoftwo
 \fi
}%
\providecommand \natexlab [1]{#1}%
\providecommand \enquote  [1]{``#1''}%
\providecommand \bibnamefont  [1]{#1}%
\providecommand \bibfnamefont [1]{#1}%
\providecommand \citenamefont [1]{#1}%
\providecommand \href@noop [0]{\@secondoftwo}%
\providecommand \href [0]{\begingroup \@sanitize@url \@href}%
\providecommand \@href[1]{\@@startlink{#1}\@@href}%
\providecommand \@@href[1]{\endgroup#1\@@endlink}%
\providecommand \@sanitize@url [0]{\catcode `\\12\catcode `\$12\catcode
  `\&12\catcode `\#12\catcode `\^12\catcode `\_12\catcode `\%12\relax}%
\providecommand \@@startlink[1]{}%
\providecommand \@@endlink[0]{}%
\providecommand \url  [0]{\begingroup\@sanitize@url \@url }%
\providecommand \@url [1]{\endgroup\@href {#1}{\urlprefix }}%
\providecommand \urlprefix  [0]{URL }%
\providecommand \Eprint [0]{\href }%
\providecommand \doibase [0]{http://dx.doi.org/}%
\providecommand \selectlanguage [0]{\@gobble}%
\providecommand \bibinfo  [0]{\@secondoftwo}%
\providecommand \bibfield  [0]{\@secondoftwo}%
\providecommand \translation [1]{[#1]}%
\providecommand \BibitemOpen [0]{}%
\providecommand \bibitemStop [0]{}%
\providecommand \bibitemNoStop [0]{.\EOS\space}%
\providecommand \EOS [0]{\spacefactor3000\relax}%
\providecommand \BibitemShut  [1]{\csname bibitem#1\endcsname}%
\let\auto@bib@innerbib\@empty
\bibitem [{\citenamefont {{Blatt}}(2007)}]{blatt2}%
  \BibitemOpen
  \bibfield  {author} {\bibinfo {author} {\bibfnamefont {R.}~\bibnamefont
  {{Blatt}}},\ }in\ \href@noop {} {\emph {\bibinfo {booktitle} {Proceedings of
  the XVIII International Conference on Laser Science (ICOLS)}}}\ (\bibinfo
  {year} {2007})\ pp.\ \bibinfo {pages} {207--215}\BibitemShut {NoStop}%
\bibitem [{\citenamefont {{Gulde}}\ \emph {et~al.}(2003)\citenamefont
  {{Gulde}}, \citenamefont {{Riebe}}, \citenamefont {{Lancaster}},
  \citenamefont {{Becher}}, \citenamefont {{Eschner}}, \citenamefont
  {{H\"affner}}, \citenamefont {{Schmidt-Kaler}}, \citenamefont {{Chuang}},\
  and\ \citenamefont {{Blatt}}}]{gulde2}%
  \BibitemOpen
  \bibfield  {author} {\bibinfo {author} {\bibfnamefont {S.}~\bibnamefont
  {{Gulde}}}, \bibinfo {author} {\bibfnamefont {M.}~\bibnamefont {{Riebe}}},
  \bibinfo {author} {\bibfnamefont {G.~P.~T.}\ \bibnamefont {{Lancaster}}},
  \bibinfo {author} {\bibfnamefont {C.}~\bibnamefont {{Becher}}}, \bibinfo
  {author} {\bibfnamefont {J.}~\bibnamefont {{Eschner}}}, \bibinfo {author}
  {\bibfnamefont {H.}~\bibnamefont {{H\"affner}}}, \bibinfo {author}
  {\bibfnamefont {F.}~\bibnamefont {{Schmidt-Kaler}}}, \bibinfo {author}
  {\bibfnamefont {I.~L.}\ \bibnamefont {{Chuang}}}, \ and\ \bibinfo {author}
  {\bibfnamefont {R.}~\bibnamefont {{Blatt}}},\ }\href@noop {} {\bibfield
  {journal} {\bibinfo  {journal} {Nature}\ }\textbf {\bibinfo {volume} {421}},\
  \bibinfo {pages} {48} (\bibinfo {year} {2003})}\BibitemShut {NoStop}%
\bibitem [{\citenamefont {Chiaverini}\ \emph {et~al.}(2005)\citenamefont
  {Chiaverini}, \citenamefont {Britton}, \citenamefont {Leibfried},
  \citenamefont {Knill}, \citenamefont {Barrett}, \citenamefont {Blakestad},
  \citenamefont {Itano}, \citenamefont {Jost}, \citenamefont {Langer},
  \citenamefont {Ozeri}, \citenamefont {Schaetz},\ and\ \citenamefont
  {Wineland}}]{chiaverini1}%
  \BibitemOpen
  \bibfield  {author} {\bibinfo {author} {\bibfnamefont {J.}~\bibnamefont
  {Chiaverini}}, \bibinfo {author} {\bibfnamefont {J.}~\bibnamefont {Britton}},
  \bibinfo {author} {\bibfnamefont {D.}~\bibnamefont {Leibfried}}, \bibinfo
  {author} {\bibfnamefont {E.}~\bibnamefont {Knill}}, \bibinfo {author}
  {\bibfnamefont {M.~D.}\ \bibnamefont {Barrett}}, \bibinfo {author}
  {\bibfnamefont {R.~B.}\ \bibnamefont {Blakestad}}, \bibinfo {author}
  {\bibfnamefont {W.~M.}\ \bibnamefont {Itano}}, \bibinfo {author}
  {\bibfnamefont {J.~D.}\ \bibnamefont {Jost}}, \bibinfo {author}
  {\bibfnamefont {C.}~\bibnamefont {Langer}}, \bibinfo {author} {\bibfnamefont
  {R.}~\bibnamefont {Ozeri}}, \bibinfo {author} {\bibfnamefont
  {T.}~\bibnamefont {Schaetz}}, \ and\ \bibinfo {author} {\bibfnamefont
  {D.~J.}\ \bibnamefont {Wineland}},\ }\href@noop {} {\bibfield  {journal}
  {\bibinfo  {journal} {Science}\ }\textbf {\bibinfo {volume} {308}},\ \bibinfo
  {pages} {997} (\bibinfo {year} {2005})}\BibitemShut {NoStop}%
\bibitem [{\citenamefont {{Brickman}}\ \emph {et~al.}(2005)\citenamefont
  {{Brickman}}, \citenamefont {{Haljan}}, \citenamefont {{Lee}}, \citenamefont
  {{Acton}}, \citenamefont {{Deslauriers}},\ and\ \citenamefont
  {{Monroe}}}]{brickman}%
  \BibitemOpen
  \bibfield  {author} {\bibinfo {author} {\bibfnamefont {K.}~\bibnamefont
  {{Brickman}}}, \bibinfo {author} {\bibfnamefont {P.~C.}\ \bibnamefont
  {{Haljan}}}, \bibinfo {author} {\bibfnamefont {P.~J.}\ \bibnamefont {{Lee}}},
  \bibinfo {author} {\bibfnamefont {M.}~\bibnamefont {{Acton}}}, \bibinfo
  {author} {\bibfnamefont {L.}~\bibnamefont {{Deslauriers}}}, \ and\ \bibinfo
  {author} {\bibfnamefont {C.}~\bibnamefont {{Monroe}}},\ }\href@noop {}
  {\bibfield  {journal} {\bibinfo  {journal} {Phys. Rev. A}\ }\textbf {\bibinfo
  {volume} {72}},\ \bibinfo {pages} {050306} (\bibinfo {year}
  {2005})}\BibitemShut {NoStop}%
\bibitem [{\citenamefont {{Lanyon}}\ \emph {et~al.}(2011)\citenamefont
  {{Lanyon}}, \citenamefont {{Hempel}}, \citenamefont {{Nigg}}, \citenamefont
  {{M\"uller}}, \citenamefont {{Gerritsma}}, \citenamefont {{Z\"ahringer}},
  \citenamefont {{Schindler}}, \citenamefont {{Barreiro}}, \citenamefont
  {{Rambach}}, \citenamefont {{Kirchmair}}, \citenamefont {{Hennrich}},
  \citenamefont {{Zoller}}, \citenamefont {{Blatt}},\ and\ \citenamefont
  {{Roos}}}]{lanyon2011}%
  \BibitemOpen
  \bibfield  {author} {\bibinfo {author} {\bibfnamefont {B.~P.}\ \bibnamefont
  {{Lanyon}}}, \bibinfo {author} {\bibfnamefont {C.}~\bibnamefont {{Hempel}}},
  \bibinfo {author} {\bibfnamefont {D.}~\bibnamefont {{Nigg}}}, \bibinfo
  {author} {\bibfnamefont {M.}~\bibnamefont {{M\"uller}}}, \bibinfo {author}
  {\bibfnamefont {R.}~\bibnamefont {{Gerritsma}}}, \bibinfo {author}
  {\bibfnamefont {F.}~\bibnamefont {{Z\"ahringer}}}, \bibinfo {author}
  {\bibfnamefont {P.}~\bibnamefont {{Schindler}}}, \bibinfo {author}
  {\bibfnamefont {J.~T.}\ \bibnamefont {{Barreiro}}}, \bibinfo {author}
  {\bibfnamefont {M.}~\bibnamefont {{Rambach}}}, \bibinfo {author}
  {\bibfnamefont {G.}~\bibnamefont {{Kirchmair}}}, \bibinfo {author}
  {\bibfnamefont {M.}~\bibnamefont {{Hennrich}}}, \bibinfo {author}
  {\bibfnamefont {P.}~\bibnamefont {{Zoller}}}, \bibinfo {author}
  {\bibfnamefont {R.}~\bibnamefont {{Blatt}}}, \ and\ \bibinfo {author}
  {\bibfnamefont {C.~F.}\ \bibnamefont {{Roos}}},\ }\href@noop {} {\bibfield
  {journal} {\bibinfo  {journal} {Science}\ } (\bibinfo {year}
  {2011})}\BibitemShut {NoStop}%
\bibitem [{\citenamefont {{\protect Army Research Office
  (USA)}}(2004)}]{roadmap}%
  \BibitemOpen
  \bibfield  {author} {\bibinfo {author} {\bibnamefont {{\protect Army Research
  Office (USA)}}},\ }\href@noop {} {\enquote {\bibinfo {title} {Arda quantum
  information science and technology roadmap:
  \texttt{http://qist.lanl.gov/qcomp\_map.shtml}}}}, (\bibinfo {year}
  {2004})\BibitemShut {NoStop}%
\bibitem [{\citenamefont {{Steane}}(2007)}]{steane2}%
  \BibitemOpen
  \bibfield  {author} {\bibinfo {author} {\bibfnamefont {A.~M.}\ \bibnamefont
  {{Steane}}},\ }\href@noop {} {\bibfield  {journal} {\bibinfo  {journal}
  {Quantum Information and Computation}\ }\textbf {\bibinfo {volume} {7}},\
  \bibinfo {pages} {171} (\bibinfo {year} {2007})}\BibitemShut {NoStop}%
\bibitem [{\citenamefont {Kielpinski}, \citenamefont {Monroe},\ and\
  \citenamefont {Wineland}(2002)}]{kielpinski}%
  \BibitemOpen
  \bibfield  {author} {\bibinfo {author} {\bibfnamefont {D.}~\bibnamefont
  {Kielpinski}}, \bibinfo {author} {\bibfnamefont {C.}~\bibnamefont {Monroe}},
  \ and\ \bibinfo {author} {\bibfnamefont {D.~J.}\ \bibnamefont {Wineland}},\
  }\href@noop {} {\bibfield  {journal} {\bibinfo  {journal} {Nature}\ }\textbf
  {\bibinfo {volume} {417}},\ \bibinfo {pages} {709} (\bibinfo {year}
  {2002})}\BibitemShut {NoStop}%
\bibitem [{\citenamefont {Schaetz}\ \emph {et~al.}(2004)\citenamefont
  {Schaetz}, \citenamefont {Leibfried}, \citenamefont {Chiaverini},
  \citenamefont {Barrett}, \citenamefont {Britton}, \citenamefont {DeMarco},
  \citenamefont {Itano}, \citenamefont {Jost}, \citenamefont {Langer},\ and\
  \citenamefont {Wineland}}]{schaetz}%
  \BibitemOpen
  \bibfield  {author} {\bibinfo {author} {\bibfnamefont {T.}~\bibnamefont
  {Schaetz}}, \bibinfo {author} {\bibfnamefont {D.}~\bibnamefont {Leibfried}},
  \bibinfo {author} {\bibfnamefont {J.}~\bibnamefont {Chiaverini}}, \bibinfo
  {author} {\bibfnamefont {M.}~\bibnamefont {Barrett}}, \bibinfo {author}
  {\bibfnamefont {J.}~\bibnamefont {Britton}}, \bibinfo {author} {\bibfnamefont
  {B.}~\bibnamefont {DeMarco}}, \bibinfo {author} {\bibfnamefont
  {W.}~\bibnamefont {Itano}}, \bibinfo {author} {\bibfnamefont
  {J.}~\bibnamefont {Jost}}, \bibinfo {author} {\bibfnamefont {C.}~\bibnamefont
  {Langer}}, \ and\ \bibinfo {author} {\bibfnamefont {D.}~\bibnamefont
  {Wineland}},\ }\href@noop {} {\bibfield  {journal} {\bibinfo  {journal}
  {Applied Physics B: Lasers and Optics}\ }\textbf {\bibinfo {volume} {79}},\
  \bibinfo {pages} {979} (\bibinfo {year} {2004})}\BibitemShut {NoStop}%
\bibitem [{\citenamefont {{Chiaverini}}\ \emph {et~al.}(2005)\citenamefont
  {{Chiaverini}}, \citenamefont {{Blakestad}}, \citenamefont {{Britton}},
  \citenamefont {{Jost}}, \citenamefont {{Langer}}, \citenamefont
  {{Leibfried}}, \citenamefont {{Ozeri}},\ and\ \citenamefont
  {{Wineland}}}]{chiaverini2}%
  \BibitemOpen
  \bibfield  {author} {\bibinfo {author} {\bibfnamefont {J.}~\bibnamefont
  {{Chiaverini}}}, \bibinfo {author} {\bibfnamefont {R.~B.}\ \bibnamefont
  {{Blakestad}}}, \bibinfo {author} {\bibfnamefont {J.}~\bibnamefont
  {{Britton}}}, \bibinfo {author} {\bibfnamefont {J.~D.}\ \bibnamefont
  {{Jost}}}, \bibinfo {author} {\bibfnamefont {C.}~\bibnamefont {{Langer}}},
  \bibinfo {author} {\bibfnamefont {D.}~\bibnamefont {{Leibfried}}}, \bibinfo
  {author} {\bibfnamefont {R.}~\bibnamefont {{Ozeri}}}, \ and\ \bibinfo
  {author} {\bibfnamefont {D.~J.}\ \bibnamefont {{Wineland}}},\ }\href@noop {}
  {\bibfield  {journal} {\bibinfo  {journal} {QIC}\ }\textbf {\bibinfo {volume}
  {5}},\ \bibinfo {pages} {419} (\bibinfo {year} {2005})}\BibitemShut {NoStop}%
\bibitem [{\citenamefont {Seidelin}\ \emph {et~al.}(2006)\citenamefont
  {Seidelin}, \citenamefont {Chiaverini}, \citenamefont {Reichle},
  \citenamefont {Bollinger}, \citenamefont {Leibfried}, \citenamefont
  {Britton}, \citenamefont {Wesenberg}, \citenamefont {Blakestad},
  \citenamefont {Epstein}, \citenamefont {Hume}, \citenamefont {Itano},
  \citenamefont {Jost}, \citenamefont {Langer}, \citenamefont {Ozeri},
  \citenamefont {Shiga},\ and\ \citenamefont {Wineland}}]{seidelin}%
  \BibitemOpen
  \bibfield  {author} {\bibinfo {author} {\bibfnamefont {S.}~\bibnamefont
  {Seidelin}}, \bibinfo {author} {\bibfnamefont {J.}~\bibnamefont
  {Chiaverini}}, \bibinfo {author} {\bibfnamefont {R.}~\bibnamefont {Reichle}},
  \bibinfo {author} {\bibfnamefont {J.~J.}\ \bibnamefont {Bollinger}}, \bibinfo
  {author} {\bibfnamefont {D.}~\bibnamefont {Leibfried}}, \bibinfo {author}
  {\bibfnamefont {J.}~\bibnamefont {Britton}}, \bibinfo {author} {\bibfnamefont
  {J.~H.}\ \bibnamefont {Wesenberg}}, \bibinfo {author} {\bibfnamefont {R.~B.}\
  \bibnamefont {Blakestad}}, \bibinfo {author} {\bibfnamefont {R.~J.}\
  \bibnamefont {Epstein}}, \bibinfo {author} {\bibfnamefont {D.~B.}\
  \bibnamefont {Hume}}, \bibinfo {author} {\bibfnamefont {W.~M.}\ \bibnamefont
  {Itano}}, \bibinfo {author} {\bibfnamefont {J.~D.}\ \bibnamefont {Jost}},
  \bibinfo {author} {\bibfnamefont {C.}~\bibnamefont {Langer}}, \bibinfo
  {author} {\bibfnamefont {R.}~\bibnamefont {Ozeri}}, \bibinfo {author}
  {\bibfnamefont {N.}~\bibnamefont {Shiga}}, \ and\ \bibinfo {author}
  {\bibfnamefont {D.~J.}\ \bibnamefont {Wineland}},\ }\href {\doibase
  10.1103/PhysRevLett.96.253003} {\bibfield  {journal} {\bibinfo  {journal}
  {Phys. Rev. Lett.}\ }\textbf {\bibinfo {volume} {96}},\ \bibinfo {pages}
  {253003} (\bibinfo {year} {2006})}\BibitemShut {NoStop}%
\bibitem [{\citenamefont {Turchette}\ \emph {et~al.}(2000)\citenamefont
  {Turchette}, \citenamefont {Kielpinski}, \citenamefont {King}, \citenamefont
  {Leibfried}, \citenamefont {Meekhof}, \citenamefont {Myatt}, \citenamefont
  {Rowe}, \citenamefont {Sackett}, \citenamefont {Wood}, \citenamefont {Itano},
  \citenamefont {Monroe},\ and\ \citenamefont {Wineland}}]{turchette}%
  \BibitemOpen
  \bibfield  {author} {\bibinfo {author} {\bibfnamefont {Q.~A.}\ \bibnamefont
  {Turchette}}, \bibinfo {author} {\bibfnamefont {D.}~\bibnamefont
  {Kielpinski}}, \bibinfo {author} {\bibfnamefont {B.~E.}\ \bibnamefont
  {King}}, \bibinfo {author} {\bibfnamefont {D.}~\bibnamefont {Leibfried}},
  \bibinfo {author} {\bibfnamefont {D.~M.}\ \bibnamefont {Meekhof}}, \bibinfo
  {author} {\bibfnamefont {C.~J.}\ \bibnamefont {Myatt}}, \bibinfo {author}
  {\bibfnamefont {M.~A.}\ \bibnamefont {Rowe}}, \bibinfo {author}
  {\bibfnamefont {C.~A.}\ \bibnamefont {Sackett}}, \bibinfo {author}
  {\bibfnamefont {C.~S.}\ \bibnamefont {Wood}}, \bibinfo {author}
  {\bibfnamefont {W.~M.}\ \bibnamefont {Itano}}, \bibinfo {author}
  {\bibfnamefont {C.}~\bibnamefont {Monroe}}, \ and\ \bibinfo {author}
  {\bibfnamefont {D.~J.}\ \bibnamefont {Wineland}},\ }\href {\doibase
  10.1103/PhysRevA.61.063418} {\bibfield  {journal} {\bibinfo  {journal} {Phys.
  Rev. A}\ }\textbf {\bibinfo {volume} {61}},\ \bibinfo {pages} {063418}
  (\bibinfo {year} {2000})}\BibitemShut {NoStop}%
\bibitem [{\citenamefont {Labaziewicz}\ \emph {et~al.}(2008)\citenamefont
  {Labaziewicz}, \citenamefont {Ge}, \citenamefont {Antohi}, \citenamefont
  {Leibrandt}, \citenamefont {Brown},\ and\ \citenamefont
  {Chuang}}]{labaziewicz1}%
  \BibitemOpen
  \bibfield  {author} {\bibinfo {author} {\bibfnamefont {J.}~\bibnamefont
  {Labaziewicz}}, \bibinfo {author} {\bibfnamefont {Y.}~\bibnamefont {Ge}},
  \bibinfo {author} {\bibfnamefont {P.}~\bibnamefont {Antohi}}, \bibinfo
  {author} {\bibfnamefont {D.}~\bibnamefont {Leibrandt}}, \bibinfo {author}
  {\bibfnamefont {K.~R.}\ \bibnamefont {Brown}}, \ and\ \bibinfo {author}
  {\bibfnamefont {I.~L.}\ \bibnamefont {Chuang}},\ }\href@noop {} {\bibfield
  {journal} {\bibinfo  {journal} {Phys. Rev. Lett.}\ }\textbf {\bibinfo
  {volume} {100}},\ \bibinfo {pages} {013001} (\bibinfo {year}
  {2008})}\BibitemShut {NoStop}%
\bibitem [{\citenamefont {MacAlpine}\ and\ \citenamefont
  {Schildknecht}(1959)}]{macalpine}%
  \BibitemOpen
  \bibfield  {author} {\bibinfo {author} {\bibfnamefont {W.}~\bibnamefont
  {MacAlpine}}\ and\ \bibinfo {author} {\bibfnamefont {R.}~\bibnamefont
  {Schildknecht}},\ }\href@noop {} {\bibfield  {journal} {\bibinfo  {journal}
  {Proceedings of the IRE}\ }\textbf {\bibinfo {volume} {47}},\ \bibinfo
  {pages} {2099} (\bibinfo {year} {1959})}\BibitemShut {NoStop}%
\bibitem [{\citenamefont {Poitzsch}(1996)}]{poitzsch}%
\BibitemOpen
  \bibfield  {author}{
    \bibinfo {author} {\bibfnamefont {M.~E.}~\bibnamefont {Poitzsch}}\ and\ 
    \bibinfo {author} {\bibfnamefont {J.~C.}~\bibnamefont {Bergquist}}\ and\ 
    \bibinfo {author} {\bibfnamefont {W.~M.}~\bibnamefont {Itano}}\ and\ 
    \bibinfo {author} {\bibfnamefont {D.~J.}~\bibnamefont {Wineland}},} \href@noop {} {\bibfield  {journal} {
    \bibinfo  {journal} {Rev. Sci. Instrum.}\ }\textbf {\bibinfo {volume} {67}},\ 
    \bibinfo {pages} {129} (\bibinfo {year} {1996})}
\BibitemShut {NoStop}%
\bibitem [{\citenamefont {{Kumph}}, \citenamefont {{Brownnutt}},\ and\
  \citenamefont {{Blatt}}(2011)}]{kumph}%
  \BibitemOpen
  \bibfield  {author} {\bibinfo {author} {\bibfnamefont {M.}~\bibnamefont
  {{Kumph}}}, \bibinfo {author} {\bibfnamefont {M.}~\bibnamefont
  {{Brownnutt}}}, \ and\ \bibinfo {author} {\bibfnamefont {R.}~\bibnamefont
  {{Blatt}}},\ }\href@noop {} {\bibfield  {journal} {\bibinfo  {journal} {New
  Journal of Physics}\ }\textbf {\bibinfo {volume} {13}},\ \bibinfo {pages}
  {073043} (\bibinfo {year} {2011})}\BibitemShut {NoStop}%
\bibitem [{\citenamefont {Abrie}(1999)}]{abrie}%
  \BibitemOpen
  \bibfield  {author} {\bibinfo {author} {\bibfnamefont {P.~L.~D.}\
  \bibnamefont {Abrie}},\ }\href@noop {} {\emph {\bibinfo {title} {Design of RF
  and Microwave Amplifiers and Oscillators}}}\ (\bibinfo  {publisher} {Artech
  House Publishers},\ \bibinfo {year} {1999})\BibitemShut {NoStop}%
\bibitem [{Note1()}]{Note1}%
  \BibitemOpen
  \bibinfo {note} {A good inductance-calculator software can be found at
  \protect \texttt {http://hamwaves.com/antennas/inductance.html}. It relies on
  several geometric and empirical correction factors, as given in \cite
  {RFcoils, inductanceformula, groverinductance}. This applet agreed with our
  experimental inductance measurements within $10\%$. In addition it can
  calculate the self-resonance frequency and an estimation of the inductor's
  quality factor.}\BibitemShut {Stop}%
\bibitem [{\citenamefont {Corum}\ and\ \citenamefont {Corum}(2001)}]{RFcoils}%
  \BibitemOpen
  \bibfield  {author} {\bibinfo {author} {\bibfnamefont {K.~L.}\ \bibnamefont
  {Corum}}\ and\ \bibinfo {author} {\bibfnamefont {J.~F.}\ \bibnamefont
  {Corum}},\ }\href@noop {} {\bibfield  {journal} {\bibinfo  {journal}
  {Microwave Review, IEEE}\ }\textbf {\bibinfo {volume} {7}},\ \bibinfo {pages}
  {36} (\bibinfo {year} {2001})}\BibitemShut {NoStop}%
\bibitem [{\citenamefont {Lundin}(1985)}]{inductanceformula}%
  \BibitemOpen
  \bibfield  {author} {\bibinfo {author} {\bibfnamefont {R.}~\bibnamefont
  {Lundin}},\ }\href@noop {} {\bibfield  {journal} {\bibinfo  {journal}
  {Proceedings of the IEEE}\ }\textbf {\bibinfo {volume} {73}},\ \bibinfo
  {pages} {1428} (\bibinfo {year} {1985})}\BibitemShut {NoStop}%
\bibitem [{\citenamefont {Grover}(2004)}]{groverinductance}%
  \BibitemOpen
  \bibfield  {author} {\bibinfo {author} {\bibfnamefont {F.}~\bibnamefont
  {Grover}},\ }\href@noop {} {\emph {\bibinfo {title} {Inductance calculations:
  working formulas and tables}}},\ Phoenix Edition Series\ (\bibinfo
  {publisher} {Dover Publications},\ \bibinfo {year} {2004})\BibitemShut
  {NoStop}%
\end{thebibliography}
\end{document}